\documentclass[eqsecnum,showpacs,showkeys,prd]{revtex4}
\usepackage{amsmath}
\begin{document}

\title{Pioneer Anomaly: An Interesting Numerical Coincidence}

\author{Jarmo M\"akel\"a}
\email[Electronic address: ]{jarmo.makela@puv.fi}
\affiliation{Vaasa University of Applied Sciences, Wolffintie 30, 65200 Vaasa, Finland}

\begin{abstract}

We note that if we construct from the observed anomalous acceleration 
$a_P = (8.74 \pm 1.33)\times 10^{-10} m/s^2$ of the spacecrafts Pioneer 10 and 11
towards the Sun, the proton mass $m_p$, and Newton's gravitational constant $G$ a 
quantity $\ell_P := (Gm_p/a_P)^{1/2}$, which has the same dimension as length, then
$\ell_P$ is roughly of the same order of magnitude as is the Compton wave length of a 
proton. We formulate a simple quantum mechanical hypothesis, which aims to provide
an explanation to this interesting numerical coincidence.

\end{abstract}



\maketitle

\section{Introduction}

    It is a pleasure and a priviledge to be a physicist nowadays. We are living an era of new
 and interesting observations which will be likely to boost a profound change in our 
physical world view. Quite a few of these new observations are related to gravitation and 
astrophysics. In short, the new observational results suggest that our current theories of
gravitation may 
need a considerable revision, especially as it comes to the properties of very weak 
gravitational fields. So far it has been generally believed that the good old Newtonian 
theory of gravity would provide an adequate description of the properties of gravity, 
when the gravitational interaction is very weak, and the speeds of the gravitating bodies
very low. After all, Newton's theory of gravitation is the weak field limit of Einstein's
general theory of relativity. According to some current observations, however, this may 
not be the case. Bodies in very weak gravitational fields seem to behave in a way
different from the one predicted by Newton's universal law of gravitation.

    One of the most interesting new observations is the so called {\it Pioneer anomaly} \cite{yksi, kaksi}. 
The spacecrafts Pioneer 10 and 11 were launched in the years 1972 and 1973, respectively,
to probe the outer reaches of our solar system. Ever since their trajectories have 
been tracked with a very great precision. Improved observational techniques, together 
with a detailed analysis of the observational data, revealed during the 90's that, in
addition to the gravitational acceleration predicted by Newton's universal law of 
gravitation, those spacecrafts possess a certain unexplained, anomalous acceleration
towards the Sun. Nowadays there is a general agreement that the magnitude of this anomalous
acceleration is \cite{yksi}
\begin{equation}
a_P = (8.74 \pm 1.33)\times 10^{-10}m/s^2.
\end{equation}
Although this acceleration is very small -it is about the same as is the gravitational      
acceleration caused by a 10 kg mass at the 1 meter distance- there is no doubt that the 
Pioneer anomaly is real. So far the Pioneer anomaly has escaped all attempted explanations 
based on the known physics. The most plausible explanation offered so far has been an 
asymmetric leakage of heat from the spacecraft \cite{kolme}. Other attempted explanations include, among
other things, the drag of interplanetary dust \cite{nelja}, and the gravitational attraction of the
objects of the Kuiper belt \cite{viisi}. All these explanations, except the asymmetric leakage of heat, 
which is still under scrutiny, however, have failed in the close examination. Even the 
attempts to explain the Pioneer anomaly by means of modifications of the current theories 
of gravity meet with a considerable difficulty: Observations made on the motions of the 
outer planets strongly suggest that the planets do not possess any anomalous accelerations but 
they move strictly according to the laws of classical physics \cite{kuusi}. So if we want to explain 
the Pioneer anomaly by means of a new theory of gravity, that theory should predict new
effects for the motions of objects whose masses are of the order of 100 kg, which is the 
order of magnitude of the masses of the spacecrafts Pioneer 10 and 11, and leave intact 
the predictions of the Newtonian theory of gravity as it comes to the motions of the
planetary objects. A very great challenge, indeed! 

   The purpose of this paper is to sketch a new type of explanation for the Pioneer anomaly.
Our starting point is a certain curious numerical coincidence: If we construct from the 
anomalous acceleration $a_P$ of Eq.(1.1), Newton's gravitational constant 
$G\approx 6.673\times 10^{-11}Nm^2/kg^2$, and the proton rest mass 
$m_p\approx 1.6725\times 10^{-27}kg$ a quantity
\begin{equation}
\ell_P := \sqrt{\frac{Gm_p}{a_P}},
\end{equation}
which has the same dimension as length, we get:
\begin{equation}
\ell_P = (1.13 \pm 0.10)\times 10^{-14}m.
\end{equation}
Curiously, this quantity, which we shall christen as the "Pioneer length" is roughly of
the same order of magnitude as is the Compton wave length of the proton:
\begin{equation}
\lambda_C := \frac{h}{m_p c} \approx 1.321\times 10^{-15}m.
\end{equation}
Indeed, the quantities $\ell_P$ and $\lambda_C$ differ from each other only by a factor 
of nine.

  The fact that the quantities $\ell_P$ and $\lambda_C$ are so surprisingly close to
each other naturally begs for an explanation, especially since the mass of a spacecraft 
consists, almost exclusively, of the masses of its protons and neutrons, which are about 
the same. One may view the relationship between $\ell_P$ and $\lambda_C$ as a relationship
between the anomalous acceleration $a_P$ and the proton mass $m_p$. It is extremely 
exciting that this relationship between $a_P$ and $m_p$ involves the Planck constant $h$. An
unexpected appearance of $h$ in this relationship makes an idea that quantum effects may 
play some role in the Pioneer anomaly almost impossible to resist. Indeed, if the Pioneer 
anomaly does not find a more mundane explanation within the context of classical physics 
during the next few years, one might almost begin to feel tempted to suspect, whether the
Pioneer anomaly might be a consequence of some subtle quantum effect of gravity.

   The basic idea of this paper is to approach the Pioneer anomaly by means of a concept
which we shall call, for the sake of brevity and simplicity, an {\it acceleration surface}.
Loosely speaking, acceleration surface may be defined as a spacelike two-surface of 
spacetime accelerating uniformly to the direction of its normal. The simplest
possible example of an acceleration surface is a flat spacelike two-plane in flat Minkowski 
spacetime accelerating to the direction of its normal. Other examples include, among other things, a 
spherical two-surface with a constant Schwarzschild coordinate $r$ in Schwarzschild 
spacetime. 
 
   When matter flows through an acceleration surface, it interacts with the geometry of the
surface. As a result, the area of the acceleration surface will change in a certain manner.
The area change as a function of the proper time $\tau$ measured by an observer at rest with 
respect to the acceleration surface depends on the energy momentum  stress tensor 
$T_{\mu\nu}$ of the matter, and may be calculated by using Einstein's field equation. Quite
recently, it has been shown that the converse is also true: If one {\it assumes} that
the area of an acceleration surface depends, as a function of the proper time $\tau$, on
the tensor $T_{\mu\nu}$ in a a certain manner, one may {\it derive} Einstein's field 
equation \cite{seitseman, kahdeksan}. In other words, one may reduce Einstein's general relativity with all
of its consequences to the properties of acceleration surfaces. The idea that one could
reduce classical general relativity to the behavior of certain spacelike two-surfaces when 
matter flows through those surfaces is far from new: It was shown already in 1995 by 
Jacobson that one may obtain Einstein's field equation by assuming that when matter
flows through a finite part of a local Rindler horizon of an accelerating observer, then 
that part shrinks such that the amount of entropy carried by matter through that part is,
in natural units, exactly one-quarter of the decrease in its area \cite{yhdeksan}. In a sense, the result 
that one may obtain classical general relativity from the assumed properties of acceleration
surfaces is just an outgrowth and a generalization of Jacobson's results.

  The fact that classical general relativity may be reduced to the properties of 
acceleration surfaces gives a rise to an idea to use those properties as a new, fresh
starting point for the attempts to quantize gravity. In these attempts the Pioneer 
anomaly might provide useful observational guidance. The acceleration surfaces presumably
have some quantum mechanical properties, and these properties should explain, among other
things, the Pioneer anomaly.

   In this paper we formulate a certain general hypothesis concerning the quantum 
mechanical properties of acceleration surfaces. Our hypothesis implies, among other
things, that protons and neutrons gain, outside of the influence of the heavy planets, 
towards the Sun an anomalous acceleration, which is about the same as is the acceleration 
$a_P$ of Eq.(1.1). Since the masses of the spacecrafts Pioneer 10 and 11 consist almost
exclusively of the masses of their protons and neutrons, our hypothesis provides an 
explanation to the Pioneer anomaly: The quantum mechanical properties of acceleration 
surfaces give a certain acceleration for protons and neutrons, and since spacecrafts are
made of protons and neutrons, they get that same acceleration.

  We begin our discussion in Section 2 by defining properly the concept of acceleration 
surface, and we consider its properties. Among other things, we shall show that for an 
acceleration surface it is possible to define, from the point of view of an observer at 
rest with respect to the surface, a quantity which we shall call as the {\it heat change}
of the surface, and which is proportional to the change in its area. It turns out that this
quantity has an interesting property that if the matter initially at rest with respect
to an acceleration surface in its immediate vicinity flows through the surface from its one
to its other side, 
then the energy transported by the matter through the surface is always, at least under certain 
fairly general conditions posed for the matter fields, of the same order of magnitude as is 
the negative of the corresponding heat change of the surface. In other words, some amount
of the thermal energy of the acceleration surface has been converted to the energy of 
the matter flowing through the surface. We may use this result when we estimate the area 
change of an acceleration surface when particles of matter (protons, for example) are 
carried through the surface.  
 
   In Section 3 we formulate in details our hypothesis concerning the quantum mechanical 
properties of acceleration surfaces. To put it simply, our hypothesis states that certain 
acceleration surfaces may absorb elementary particles by means of quantum mechanical 
processes such that when an acceleration surface absorbs a particle, the resulting increase
in the area of the surface is about the same as is the area occupied by the particle on that
surface. Since the quantum mechanical "size" of an elementary particle is, in a certain 
sense, given by its Compton wave length $\lambda_C$, we may approximate the area occupied
by an elementary particle on an acceleration surface by the surface area of a sphere having 
the Compton wave length $\lambda_C$ of the particle as its radius. Our hypothesis 
immediately implies, together with the results obtained in Section 2, that the "Pioneer
length" $\ell_P$ of Eq.(1.2) is about the same as is the Compton wave length $\lambda_C$ of
the proton. Our hypothesis also tells the direction of the anomalous acceleration gained by 
elementary particles in a gravitational field.

   We close our discussion in Section 4 with some concluding remarks.

\maketitle

      \section{Acceleration Surfaces and Their Properties}

     Quite recently it has been found that Einstein's field equation, and thereby the whole 
classical gravity, may be obtained from the properties of the so called {\it acceleration
surfaces} \cite{seitseman, kahdeksan}. To put it simply, acceleration surface is a spacelike two-surface of spacetime, 
which is in a uniformly accelerating motion to the direction of its normal. More precisely,
an acceleration surface is defined as a smooth, orientable, spacelike two-surface of 
spacetime such that the proper acceleration vector field $a^\mu$ of the congruence of the
timelike world lines of its points has the property 
\begin{equation}
\sqrt{a^\mu a_\mu} = constant := a,
\end{equation}
and there exists a spacelike unit normal vector field $n^\mu$ for the surface such that
\begin{equation}
a^\mu n_\mu \equiv a.
\end{equation}
In other words, the absolute value $a$ of the proper acceleration is identically constant
everywhere and all the time on an acceleration surface. The world lines of the points of 
an acceleration surface are parametrized by the proper time $\tau$ measured along these
world lines, and this proper time gives the same time coordinate for every point on the 
surface, i.e. acceleration surfaces are specific $\tau=constant$ two-surfaces of spacetime.

    The simplest possible example of an acceleration surface is a flat, spacelike two-plane
in flat, four-dimensional Minkowski spacetime with a constant proper acceleration $a$ to the
direction of its spacelike unit normal. As another example we may consider a spacelike
two-sphere $r=constant$ in Schwarzschild spacetime equipped with a Schwarzschild metric
\begin{equation}
ds^2 = - (1 - \frac{2M}{r})\,dt^2 + \frac{dr^2}{1 - \frac{2M}{r}} + r^2(d\theta^2 
+ \sin^2\theta\,d\phi^2).
\end{equation}
The only non-zero component of the future directed unit tangent vector field $u^\mu$ of
the world lines of the two-sphere under consideration is
\begin{equation}
u^t = (1 - \frac{2M}{r})^{-1/2},
\end{equation}
and the two-sphere has a spacelike unit normal vector field $n^\mu$, whose only non-zero 
component is
\begin{equation}
n^r = (1-\frac{2M}{r})^{1/2}.
\end{equation}
Hence the only non-zero component of the proper acceleration vector field 
$a^\mu:=u^\alpha u^\mu_{;\alpha}$ is
\begin{equation}
a^r = u^tu^r_{;t} = \frac{M}{r^2},
\end{equation}
which means that all points of the two-sphere have all the time the same constant proper
acceleration 
\begin{equation}
a = \sqrt{a^\mu a_\mu} = a^\mu n_\mu = (1 - \frac{2M}{r})^{-1/2}\frac{M}{r^2}
\end{equation}
to the direction of the vector $n^\mu$. In other words, the two-sphere $r=constant$ indeed
is an acceleration surface.

    The main motivation for defining the concept of acceleration surface is that 
acceleration surfaces are very similar to the {\it event horizons of black holes}: According
to the zeroth law of black hole mechanics the surface gravity $\kappa$ is constant 
everywhere and all the time on a black hole event horizon, whereas on an acceleration 
surface the proper acceleration $a$ is constant. Moreover, black hole event horizon may 
always be regarded as as an asymptotic limit of a certain acceleration surface, when the
proper acceleration $a$ on that acceleration surface goes to infinity. For example, 
one may observe from Eq.(2.7) that in the limit, where $a\longrightarrow\infty$ on the two-
sphere $r=constant$, $r$ must approach $2M$, the Schwarzschild radius of the Schwarzschild
black hole. In other words, the acceleration surface $r=constant$ becomes to the event 
horizon of a Schwarzschild black hole in the limit, where $a\longrightarrow\infty$.

    Acceleration surfaces have certain very interesting {\it thermodynamical} properties, 
which are very similar to those of black hole event horizons. For instance, according
to a distant observer a black hole event horizon with a surface gravity $\kappa$ has the 
Hawking temperature \cite{kymmenen}
\begin{equation}
T_H := \frac{\kappa}{2\pi},
\end{equation}
whereas an observer at rest with respect to an acceleration surface observes thermal
radiation, whose temperature is the so called Unruh temperature \cite{yksitoista}
\begin{equation}
T_U := \frac{a}{2\pi}
\end{equation}
even when, according to all inertial observers, all matter fields are in vacuum. We may 
regard the Unruh temperature $T_U$ as the temperature of an acceleration surface in the 
same sense as the Hawking temperature $T_H$ may be regarded as the temperature of a black 
hole event horizon.
    
    If we accept the view that acceleration surfaces, in the same way as black hole event
horizons, have a certain temperature, we are forced to conclude that an acceleration surface
possesses, from the point of view of an observer at rest with respect to the surface, a 
certain amount of {\it heat}. As it is well known, variations of the surface gravity 
$\kappa$ and the area $A$ of a black hole event horizon have the following property 
\cite{kaksitoista}:
\begin{equation}
\delta(\frac{1}{4\pi}\kappa A) = \frac{1}{4\pi}\delta\kappa A + \frac{1}{4\pi}\kappa\delta A 
= \frac{1}{8\pi}\kappa\delta A.
\end{equation}
Since $1/4\delta A$ is the change in the entropy of the horizon, and $\kappa/(2\pi)$ is its
Hawking temperature, the thermodynamical relation $\delta Q = T\,dS$ implies that the
{\it heat change} of a black hole event horizon is
\begin{equation}
\delta Q_{bh} = \delta(\frac{1}{4\pi}\kappa A).
\end{equation}
Analogies between acceleration surfaces and black hole event horizons therefore suggest that
if the area $A$ of an acceleration surface experiences a change $\Delta A$, then the 
corresponding change in the heat content of the acceleration surface is:
\begin{equation}
\Delta Q_{as} = \Delta(\frac{1}{4\pi} aA) = \frac{1}{4\pi}a\,\Delta A
\end{equation}
or, in SI units:
\begin{equation}
\Delta Q_{as} = \frac{c^2}{4\pi G} a\,\Delta A.
\end{equation}
The last equality in Eq.(2.12) follows from the assumption that $a$ is kept constant during
the process, where the area $A$ changes. When we calculate the change $\Delta A$ in the 
area $A$ of an acceleration
surface, we follow the world lines of the points of the surface, and we parametrize the world
lines by the proper time $\tau$ measured along those world lines. By the change $\Delta A$
of the area $A$ we mean the change in the area of the surface $\tau=constant$, when $\tau$
is changed.

  There are good grounds to believe that the quantity $\Delta Q_{as}$ of Eq.(2.13) really 
describes the change in the heat content of an acceleration surface. For instance, it may
be shown that Einstein's field equation with a vanishing cosmological constant may be 
obtained for general matter fields from an equation \cite{seitseman}
\begin{equation}
\frac{\delta^2 Q_{rad}}{d\tau^2}\vert_{\tau=0} 
= -\frac{\delta^2 Q_{as}}{d\tau^2}\vert_{\tau=0},
\end{equation}
provided that
\begin{equation}
\frac{\delta Q_{as}}{d\tau}\vert_{\tau=0} = 0.
\end{equation}
In Eq.(2.14) $\frac{\delta Q_{rad}}{d\tau}$ means the flow of heat (heat flown during a unit
proper time) carried by massless, noninteracting radiation through an acceleration surface of 
spacetime, and $\frac{\delta^2 Q_{rad}}{d\tau^2}$ denotes the rate of change in this heat 
flow. Since Eq.(2.14) implies Einstein's field equation, and therefore the whole classical 
general relativity with all of its consequences, we may view Eq.(2.14) as a fundamental 
equation in the thermodynamics of spacetime. Although Eq.(2.14) has been written for 
massless, noninteracting radiation only, it may be used for general matter fields: If
the acceleration surface is assumed to move, when $\tau=0$, with a velocity very close to 
that of light with respect to the matter fields, the kinetic energies of the particles of 
the fields vastly exceed the other forms of energies, and we may consider matter, from the
point of view of an observer at rest with respect to the acceleration surface, in effect, 
as a gas of non-interacting massless particles \cite{kahdeksan}.

    It is interesting to consider the special case, where an acceleration surface is at rest 
with respect to the radiation, when $\tau=0$. This means that
\begin{equation}
\frac{\delta Q_{rad}}{d\tau}\vert_{\tau=0} = 0,
\end{equation}
and therefore we may write $\Delta Q_{rad}$, the amount of heat carried by radiation 
through the acceleration surface during a proper time interval $[0,\tau]$, as well as 
the corresponding change $\Delta Q_{as}$ in the heat content of the acceleration surface,
as a Taylor expansion:
\begin{subequations}
\begin{eqnarray}
\Delta Q_{rad} &=& \frac{1}{2}\frac{\delta^2 Q_{rad}}{d\tau^2}\vert_{\tau=0}\tau^2 
+ O(\tau^3),\\
\Delta Q_{as} &=& \frac{1}{2}\frac{\delta^2 Q_{as}}{d\tau^2}\vert_{\tau=0}\tau^2 + O(\tau^3),
\end{eqnarray}
\end{subequations}
where $O(\tau^3)$ denotes the terms, which are of the order $\tau^3$, or higher. Hence our
fundamental equation (2.14) implies that for very small $\tau$:
\begin{equation}
\Delta Q_{rad} = - \Delta Q_{as}.
\end{equation}
This result means that if our acceleration surface is originally at rest, and then begins 
to move with respect to the radiation such that radiation flows through the acceleration
surface from its one side to another, the heat gained by the other side of the acceleration
surface is exactly the heat lost by the surface. So it appears for an observer at rest 
with respect to the surface as if the surface emitted radiation such that the heat of the
acceleration surface is exactly converted to the heat of the radiation. During the process
the area of the acceleration surface experiences, according to Eq.(2.13), the change
\begin{equation}
\Delta A = -\frac{4\pi G}{c^2}\,\Delta Q_{rad}.
\end{equation}
In other words, the acceleration surface {\it shrinks}, when radiation flows through the
surface.

    One might expect that a relationship somewhat similar to Eq.(2.18) would hold even when
the matter flowing through an acceleration surface is not just massless, non-interacting
radiation, and other forms of energy, except heat (mass-energy, for instance) are carried
through the surface. More precisely, one expects that always when matter carries energy 
through an acceleration surface such that the surface is originally at rest with respect 
to the matter, the total energy $\Delta E_{matter}$ carried by the matter through the 
surface would be, although not necessarily exactly equal, at least of the same order of 
magnitude
as is the heat lost by the acceleration surface. In other words, one might expect that
\begin{equation}
\Delta E_{matter} \sim -\Delta Q_{as},
\end{equation}
regardless of what kind of matter we happen to have. This issue has been investigated in the
Appendix. It turns out that Eq.(2.20) holds at least when the spatial geometry of the 
spacetime is, as well as is the matter distribution, homogeneous and isotropic, there are no 
negative pressures, and the matter satisfies the dominant energy condition. In particular, 
it turns out that if matter consists of homogeneous, pressureless dust only, we have:
\begin{equation}
\Delta E_{matter} = -\frac{3}{2}\,\Delta Q_{as},
\end{equation}
which is consistent with Eq.(2.20). 

\maketitle

  \section{A Hypothesis}

   So far we have learned that with acceleration surfaces, which are somewhat analogous to 
the event horizons of black holes, it is possible to associate the concept of heat change. 
Classical gravity as a whole may be formulated in terms of the heat exchange between an 
acceleration surface, and the matter which flows through that surface. When an acceleration 
surface is originally at rest with respect to the matter fields, and then begins to move
with respect to the matter, the energy carried by the matter through the acceleration 
surface during a very short proper time interval is, at least under certain conditions, 
of the same order of magnitude as is the heat lost by the acceleration surface. When 
matter consists of massless, non-interacting radiation, the amount of energy carried by
the matter, and the heat lost by the acceleration surface, are exactly the same.

  Since acceleration surfaces seem to play so central role in classical gravity, one is
prompted to consider their possible role in quantum gravity. Could acceleration surfaces
provide a new, fresh starting point for the attempts to quantize gravity? In particular, 
is it possible to explain, by means of acceleration surfaces and their possible quantum
mechanical properties, the curious numerical coincidence of Eqs.(1.3) and (1.4) between 
the anomalous acceleration $a_P$ of the spacecrafts Pioneer 10 and 11 towards the Sun, and
the Compton wavelength $\lambda_C$ of the proton?

  To approach this question, consider a particle (proton, for example) lying on an 
acceleration surface, originally at rest with respect to that surface. After a very short 
elapsed proper time measured by an observer moving along with the acceleration surface the
particle has entered through the surface. According to the results of the previous Section
our acceleration surface shrinks in this process such that the resulting decrease in the 
heat content of the surface is about the same as is the energy of the particle. In other
words, it appears for an observer at rest with respect to the acceleration surface as if
the acceleration surface had {\it emitted} a particle with a certain energy such that some 
amount of the energy of the acceleration surface is converted to the energy of the particle.
If we consider acceleration surfaces as quantum mechanical objects, an emission of a 
particle by an acceleration surface must correspond to a certain transition from a one to
another quantum state of the acceleration surface. The results obtained from loop quantum 
gravity \cite{kolmetoista}, as well as from the investigations concerning the quantum mechanical properties 
of the event horizons of black holes \cite{neljatoista}, suggest that for an acceleration surface it is 
possible to define an {\it area operator}, which has a discrete spectrum. Since acceleration
surfaces shrink during emissions of particles, it is natural to think that an emission of a 
particle by an acceleration surface corresponds to a quantum mechanical transition performed
by the acceleration surface from a one to another area eigenstate.

   Now, if we really consider the processes, where a particle comes through an acceleration 
surface as quantum mechanical emission processes of particles by the acceleration surface,
we are faced with a possibility that an acceleration surface may, in addition of emitting,
also {\it absorb} particles as well. In other words, it may be possible that an 
acceleration surface "catches" a particle originally moving along a geodesic of spacetime
and, as a result, the particle begins to move along with the acceleration surface with a 
proper acceleration equal to that of the surface. This means that a particle originally 
in a free fall may suddenly gain a certain non-zero proper acceleration, even
when it has no interactions, except gravity, with the other particles of spacetime. 
Classically, such a process is impossible, and if the process described above really exists,
it must be of a quantum mechanical origin. An absorption of a particle by an acceleration 
surface is a process inverse to that of emission, and therefore an acceleration surface 
performs, during an absorption of a particle, a quantum mechanical transition from a lower
to a higher area eigenstate.

  The absorption process of a particle by an acceleration surface may provide a possible
explanation to the Pioneer anomaly: The protons and the neutrons of a spacecraft are absorbed by
an acceleration surface possessing a proper acceleration $a_P$ towards the Sun, and therefore 
the spacecraft gains an anomalous additional acceleration $a_P$. This kind of an explanation
to the Pioneer anomaly, however, gives a rise to several questions: Why are the protons 
and the
neutrons of the spacecrafts Pioneer 10 and 11 absorbed by an acceleration surface, which is 
accelerating {\it towards} the Sun? Why are they not absorbed by acceleration surfaces, 
which are in accelerating motions in other directions? For instance, why are they not 
absorbed by an acceleration surface, which is accelerating {\it outwards} from the Sun?
Moreover, why is the proper acceleration $a_P$ of the acceleration surface which captures 
the protons and the neutrons of the Pioneer spacecrafts, about $8.7\times 10^{-10}m/s^2$? 
Why is it not, say, $9.8m/s^2$? Is there any physical principle which would determine both 
the directions and the magnitudes of the proper accelerations of those acceleration 
surfaces, which absorb the given types of particles?

   We first consider the question of the direction of the proper acceleration. To this end,
we first define the concepts of {\it negative}, {\it positive} and {\it zero} acceleration
surfaces. By zero acceleration surface $\Sigma_0$ we mean an acceleration surface, where 
the proper acceleration is zero. In other words, all points of a zero acceleration 
surface are in a free fall, and thereby they move along timelike geodesics of spacetime.

   It should be clear that for any acceleration surface $\Sigma$ there exists, in an 
arbitrary moment $\tau_0$ of the proper time $\tau$ measured along the world lines of the 
points of the surface, a zero acceleration surface $\Sigma_0$ such that the points of the
acceleration surface $\Sigma$, as well as the four-velocities $u^\mu$ of those points,
coincide with those of the zero acceleration surface $\Sigma_0$: At the moment 
$\tau=\tau_0$ one just releases all points of the acceleration surface $\Sigma$ into a free
fall, and so one gets the zero acceleration surface $\Sigma_0$. For the sake of brevity and 
simplicity we say that the zero acceleration surface $\Sigma_0$ {\it matches} with the 
acceleration surface $\Sigma$, when $\tau=\tau_0$. 

   We are now prepared to define the concepts of negative and positive acceleration 
surfaces: An acceleration surface $\Sigma_{-}$ ($\Sigma_{+}$) is a negative (positive)
acceleration surface, if at any moment $\tau=\tau_0$ the surface $\Sigma_{-}$ ($\Sigma_{+}$)
and the zero acceleration surface $\Sigma_0$ matching with $\Sigma_{-}$ ($\Sigma_{+}$) have
the following properties:

  (i) The world lines of the points of the surface $\Sigma_{-}$ ($\Sigma_{+}$) never 
intersect the world lines of the points of the surface $\Sigma_0$ for any $\tau>\tau_0$.

  (ii) There exists a proper time interval $\Delta \tau>0$ such that if we pick up any
open subset $S_{-}$ ($S_{+}$), which becomes an open subset $S_0$ of $\Sigma_0$ when 
$\tau=\tau_0$, then the area $A_{-}$ ($A_{+}$) of $S_{-}$ ($S_{+}$) is smaller (greater)
than the area $A_0$ of $S_0$ for all $\tau\in(\tau_0, \tau_0 + \Delta \tau)$. 

   It should be noted that the proper time $\tau$ on the zero acceleration surface 
$\Sigma_0$ as well as on the surface $\Sigma_{-}$ ($\Sigma_{+}$) has been measured along the 
world lines of the points of the surfaces such that when the surface $\Sigma_0$ matches 
with the surface $\Sigma_{-}$ ($\Sigma_{+}$), the proper time $\tau=\tau_0$ on the both 
surfaces. In essence, our definition of negative and positive acceleration surfaces  
just says that if we pick up, in any moment $\tau_0$ of the proper time $\tau$, any part of 
a negative (positive) acceleration surface, then immediately after the moment $\tau_0$
the area of that part is smaller (greater) than it would have been, if the points of that 
part would have been released in a free fall.

  It is very easy to give examples of negative, positive and zero acceleration surfaces.
For instance, it is straightforward to show that in Schwarzschild spacetime those spacelike
two-spheres, where the radial coordinate $r$ obeys for all $r$ an equation
\begin{equation}
\ddot{r} = -\frac{M}{r^2},
\end{equation}
where the dot means proper time derivative, are in a free fall, and therefore they are 
zero acceleration surfaces. A two-sphere, whose points are in a uniformly accelerating 
motion with a constant proper acceleration in the direction of the spacelike normal of
the sphere such that $\ddot{r}<-\frac{M}{r^2}$ for all $r$, are negative acceleration
surfaces, whereas those two-spheres, where $\ddot{r}>-\frac{M}{r^2}$ for all $r$, are
positive acceleration surfaces. Indeed, if we pick up any part of a two-sphere, where 
$\ddot{r}<-\frac{M}{r^2}$ ($>-\frac{M}{r^2}$) for all $r$, then the area of that part is,
immediately after any instant $\tau_0$ of the proper time $\tau$, smaller (greater) than 
it would have been, if the points of that part would have been released in a free fall,
when $\tau=\tau_0$.

   The spacetime geometry, where the spacecrafts Pioneer 10 and 11 move is, at least as an
excellent approximation, a Schwarzschild geometry created by the mass of the Sun. Since the
spacecrafts seem to have a certain constant, anomalous acceleration $a_P$ towards the Sun,
it appears that they move along with certain negative acceleration surfaces with a proper
acceleration $a_P$. In other words, it seems as if the protons and the neutrons of the 
spacecrafts Pioneer 10 and 11 had undergone a quantum mechanical absorption process
performed by a certain negative acceleration surface. It is tempting to speculate on the 
possibility that this is a general feature of negative acceleration surfaces: Negative 
acceleration surfaces tend to absorb elementary particles. At least such a hypothesis 
would explain the direction of the anomalous acceleration experienced by the spacecrafts 
Pioneer 10 and 11.  

  Although our hypothesis seems to explain the {\it direction} of the anomalous 
acceleration $a_P$, it does not, however, explain its {\it magnitude}. It is 
natural to think that those quantum-mechanical absorption processes performed by negative
acceleration surfaces would be favored, where the corresponding increase in the area of 
the acceleration surface is roughly of the same order of magnitude as is the "surface area"
of the particle which is being absorbed. The order of magnitude in the increase in the 
area of an acceleration surface may be calculated by means of Eqs. (2.13) and (2.20) 
such that we 
substitute for $\Delta E$ the mass energy $mc^2$ of the particle with mass $m$. The 
"radius" of the particle, in turn, may be described, in a quantum mechanical sense, by 
its Compton wave length $\lambda_C = \frac{h}{mc}$. If we know the "radius" of the
particle, we may estimate its effective "surface area".

   We now condense all of the speculations we have expressed so far on the possible 
quantum mechanical properties of acceleration surfaces into the following hypothesis, 
which at the very least seems to explain the Pioneer anomaly:

 {\it Negative acceleration surfaces of spacetime tend to absorb those elementary 
particles which have the property that the surface area of a sphere with a radius 
equal to the Compton wave length of the particle is of the same order of magnitude 
as is the area increase caused by the absorption of the particle by the acceleration 
surface.}

  We have already seen, how this hypothesis explains the direction of the anomalous 
acceleration of the Pioneer spacecrafts. To see how it explains its magnitude as well,
consider Eq.(2.20). If we take $\Delta E$ to be the mass energy $mc^2$ of a particle with 
mass $m$ we find, using Eq.(2.13), that the increase in the area $A$ of a negative 
acceleration surface during an absorption of the particle is:
\begin{equation}
\Delta A \sim \frac{4\pi Gm}{a},
\end{equation}
According to our hypothesis $\Delta A$ is of the same order of magnitude as is the surface
area of a sphere with a radius equal to the Compton wave length of the particle. This 
implies:
\begin{equation}
\Delta A \sim 4\pi\lambda_C^2 = 4\pi\frac{h^2}{m^2c^2}.
\end{equation}
Comparing Eqs.(3.2) and (3.3) we therefore find:
\begin{equation}
\lambda_C \sim \sqrt{\frac{Gm}{a}}.
\end{equation}
In other words, we have found that the curious numerical relationship, which was 
discovered in the Introduction between the Compton wave length of the proton, the 
proton mass, and the anomalous acceleration of the spacecrafts Pioneer 10 and 11, is 
of general validity, provided that our hypothesis is true: Our hypothesis implies that
elementary particles with mass $m$ receive in a gravitational field an anomalous
acceleration $a$ to the direction of a spacelike normal of a negative acceleration
surface of spacetime such that Eq.(3.4) holds. According to our hypothesis this effect
is of a purely quantum mechanical origin, and it cannot be explained by means of 
classical physics.

  Using Eq.(3.4) we may obtain a very rough order-of-magnitude estimate for the anomalous
acceleration $a$ received by an elementary particle with mass $m$:
\begin{equation}
a \sim \frac{Gm^3c^2}{h^2}.
\end{equation}
If our hypothesis is true, this expression should give, up to a numerical coefficient
of the order of unity, the correct value for the anomalous acceleration $a$. To find the
precise value of that numerical coefficient, we should carry out a detailed quantum 
mechanical calculation which, unfortunately, is out of reach at the moment. Nevertheless,
it is interesting to note that if we define a quantity
\begin{equation}
a_0 := \frac{\ln 2}{16\pi}\frac{Gm^3c^2}{h^2},
\end{equation}
we get, if we substitute for $m$ the proton mass $m_p\approx 1.6725\times 10^{-27}kg$:
\begin{equation}
a_0 \approx 8.83\times 10^{-10} m/s^2,
\end{equation}
which is well within the error bars of the observed anomalous acceleration $a_P$ of the
spacecrafts Pioneer 10 and 11.

\maketitle

      \section{Concluding Remarks}

  In this paper we have found that the Pioneer anomaly may be explained by means of a very 
simple quantum mechanical hypothesis concerning the properties of the so called 
acceleration surfaces of spacetime. Loosely speaking, acceleration surface may be defined 
as a spacelike two-surface of spacetime accelerating uniformly to the 
direction of its spacelike normal. According to our hypothesis certain acceleration 
surfaces, which we called {\it negative} acceleration surfaces, absorb elementary particles
by means of a still unknown quantum mechanical process such that when a negative 
acceleration surface absorbs an elementary particle, the resulting increase in the area
of the surface is about the same as is the area occupied by the absorbed particle on 
the surface. The area occupied by an elementary particle on an acceleration surface,
in turn, may be estimated, at least as far as we are interested in mere order-of-magnitude
approximations, by the surface area of a sphere having the Compton wave length of the
particle as its radius. Our hypothesis implied, among other things, that elementary 
particles with mass $m$ gain towards the Sun a certain anomalous acceleration, which
depends on the mass $m$. Using our hypothesis we managed to find, up to a still unknown
numerical factor of order unity, an explicit expression for that acceleration. If the
numerical factor in question is chosen to be $\frac{\ln 2}{16\pi}$, and we substitute
for the mass $m$ the proton mass $m_p$, we get for the anomalous acceleration a value, 
which is well within the error bars of the observed anomalous acceleration $a_P$ of the
spacecrafts Pioneer 10 and 11 towards the Sun. Since the masses of those spacecrafts
consist mainly of the masses of their protons and neutrons, our hypothesis seems to be
capable to explain the Pioneer anomaly: The protons and the neutrons of the spacecrafts 
Pioneer 10 and 11 are absorbed by a certain acceleration surface, and therefore the 
spacecrafts get a certain anomalous acceleration towards the Sun.

   The observational data gained so far provides some reasons to believe that, in contrast
to the spacecrafts Pioneer 10 and 11, the outer planets of our solar system do not possess 
any anomalous acceleration towards the Sun, but they do move according to the well 
established laws of classical gravity. If this conclusion drawn from the observational data
is correct, it may also be explained by our hypothesis based on the concept of acceleration 
surface: The outer planets are very heavy objects. Their masses are more than $10^{25}kg$,
whereas the masses of the spacecrafts Pioneer 10 and 11 are less than $10^3kg$. Beacuse of 
that, the gravitational field created by an outer planet dominates over the gravitational field
created by the Sun very far away from the planet. A simple calculation based on Newton's 
universal law of gravitation reveals that the gravitational field of a planet with mass
$10^{25}kg$ at the distance 10 AU from the Sun dominates over the gravitational field 
created by the Sun up to the distances of several million kilometers from the planet, 
whereas the gravitational fields created by the spacecrafts Pioneer 10 and 11 are, when
compared to the gravitational field created by the Sun, almost negligible even in their 
immediate vicinities. So it appears that the negative acceleration surface which 
accelerates towards the Sun with the proper acceleration $a_P$ lies very far away from
a planet, whereas it lies in an immediate vicinity of a spacecraft. As a consequence, the 
protons and the neutrons of the spacecraft are absorbed by that acceleration surface,
whereas those of a planet are not, and so the planet moves according to the laws of 
classical gravity. In a sense, the huge gravitational field created by the planet protects
its protons and neutrons from the quantum effects occuring in the gravitational field
created by the Sun.

    Even though our hypothesis seems to be capable to explain the magnitude of the 
anomalous acceleration experienced by the spacecrafts Pioneer 10 and 11, together 
with the absence of any observed anomalous acceleration of the outer planets, it 
may also have some problems of its own. These potential problems are related to 
the {\it direction} of the anomalous acceleration. The key problem is, whether the 
requirement that elementary particles are absorbed by negative acceleration  surfaces 
only, is enough to specify that direction. In flat Minkowski spacetime, for instance,
one may construct in a neighbourhood of any point a negative acceleration surface, where
the proper acceleration vector field $a^\mu$ points to an arbitrary spacelike direction of 
spacetime: One just associates shrinking spheres with the given point. What determines 
which negative acceleration surface the elementary particle begins to follow? If our 
hypothesis is correct, elementary particles may gain a certain anomalous acceleration
in  any direction in flat spacetime. Why do we not observe in flat spacetime bodies made 
of protons and neutrons mysteriously accelerating in arbitrary directions?

    The answer to this problem lies in the {\it symmetries} of flat spacetime. Flat 
spacetime looks exactly the same in all spatial directions, and therefore the elementary 
particles in flat spacetime are absorbed by negative acceleration surfaces accelerating
in different directions with equal probabilities. Since the probability of being 
accelerated to the given direction is the same for all directions, there is, as a net 
effect, no acceleration at all in any direction. Hence we do not observe anomalous 
accelerations for bodies in flat spacetime. If spacetime is curved, however, the spatial
symmetries of flat spacetime are broken. For instance, if we look at Schwarzschild spacetime 
from a point different from the origin, we may observe that spacetime looks different in 
different directions. As a result, absorption of an elementary particle by a negative 
acceleration surface accelerating in a certain direction becomes more probable than being
absorbed by surfaces accelerating in other directions. As a net effect we may observe 
particles and bodies propagating in curved spacetime with certain anomalous accelerations.
It will be an interesting research project of the future to investigate how the 
probability distribution associated with the directions of anomalous accelerations depends 
on the large scale geometry of spacetime. The final aim of such a project is to find out, 
whether 
anomalous accelerations for bodies in an free fall could be observed even in laboratory
conditions, provided that the spacetime geometry is sufficiently asymmetric. If our 
hypothesis is correct, one expects to be able to observe for bodies in a free fall in a 
laboratory, in addition to the ordinary gravitational acceleration caused by Earth's 
gravity, an anomalous acceleration, which is of the order of $10^{-9}m/s^2$, and is
caused by the quantum effects of gravity.   
  
\maketitle

\appendix*

\section{Boost Energy Flow and Heat Change}

  In this Appendix we shall show that Eq.(2.20) holds at least when spacetime geomerty and 
the matter distribution are homegenous and isotropic in a (small) region of spacetime under
consideration, there are no negative pressures, and the matter satisfies the dominant 
energy condition.

  Our starting point is an equation
\begin{equation}
\frac{\ddot{V}}{V}\vert_{\tau=0} = 4\pi(T^0_{\,\,0}-T^1_{\,\,1} - T^2_{\,\,2} - T^3_{\,\,3}).
\end{equation}
It has been pointed out by Baez that this equation summarizes the geometric content of 
Einstein's field equation \cite{viisitoista}. In this equation $V$ denotes the three-volume of a very small
three-dimensional spatial region of spacetime. The dot denotes the proper time derivative
from the point of view of an observer in a free fall, and we have assumed that
\begin{equation}
\frac{dV}{d\tau}\vert_{\tau=0} = 0.
\end{equation}
The components of the energy momentum stress tensor $T^\mu_{\,\,\nu}$ of the matter have 
been written in an othonormal geodesic system of coordinates. In this system of coordinates
\begin{equation}
\rho := -T^0_{\,\,0}
\end{equation}
describes the energy density of the matter, whereas the quantities
\begin{equation}
p_k := T^k_{\,\,k},
\end{equation}
where $k=1,2,3$, are the pressure components. Hence we may write Eq.(A.1) as:
\begin{equation}
\frac{\ddot{V}}{V}\vert_{\tau=0} = -4\pi(\rho + p_1 + p_2 + p_3).
\end{equation}
Since the matter distribution is assumed to be homogeneous and isotropic, we have
\begin{equation}
p_1 = p_2 = p_3 := p,
\end{equation}
and the dominant energy condition, together with the non-negativity of the pressure, 
implies \cite{kaksitoista}:
\begin{equation}
\rho \geq p \geq 0.
\end{equation}
So we find that Eq.(A.1) takes finally the form:
\begin{equation}
\frac{\ddot{V}}{V}\vert_{\tau=0} = -4\pi(\rho + 3p).
\end{equation}
We may assume that matter is at rest with respect to our system of coordinates.

  At this point we note that since, in addition to the matter fields, the spacetime 
geometry is also assumed to be homogeneous and isotropic, the spatial region of spacetime
under consideration expands and contracts in the same ways in all spatial directions. So
we find that if we pick up from spacetime an acceleration surface at rest with respect to
our system of coordinates, when $\tau=0$, Eq.(A.2) implies:
\begin{equation}
\frac{dA}{d\tau}\vert_{\tau=0} = 0,
\end{equation}
where $A$ is the area of the acceleration surface. Because the area $A$ scales as the power
2/3 of the volume $V$, we observe:
\begin{equation}
\frac{\ddot A}{A}\vert_{\tau=0} = \frac{2}{3}\frac{\ddot{V}}{V}\vert_{\tau=0},
\end{equation}
and so Eq.(A.8) implies that
\begin{equation}
\frac{1}{4\pi}\frac{d^2 A}{d\tau^2} = -\frac{2}{3}A(\rho + 3p),
\end{equation}
when $\tau=0$. After a very short proper time interval $\tau$ we therefore find that the 
area of the acceleration surface has experienced the change
\begin{equation}
\Delta A = -\frac{4\pi}{3}A(\rho + 3p)\tau^2 + O(\tau^3),
\end{equation}
and the heat content of the surface the change
\begin{equation}
\Delta Q_{as} = -\frac{1}{3} aA(\rho + 3p)\tau^2 + O(\tau^3),
\end{equation}
where $a$ is the proper acceleration of our acceleration surface.

  Consider now the flow of energy through the acceleration surface. In general, the boost 
energy flow (boost energy flown during a unit time) through a very small spacelike 
two-surface with area $A$ is
\begin{equation}
\frac{dE_{matter}}{d\tau} = AT_{\mu\nu}u^\mu n^\nu,
\end{equation}
where $u^\mu$ is the future directed unit tangent vector of the world line of the surface,
and $n^\mu$ is a spacelike unit normal vector of the surface. If the surface is in a 
uniformly accelerating motion to the direction of the vector $n^\mu$ with a proper 
acceleration $a$, the vectors $u^\mu$ and $n^\mu$ transform to the vectors
\begin{subequations}
\begin{eqnarray}
u'^\mu &=& u^\mu + a\tau\,n^\mu,\\
n'^\mu &=& a\tau\, u^\mu + n^\mu,
\end{eqnarray}
\end{subequations}
during a short proper time interval $\tau$. In Eq.(A.15) we have assumed that $a\tau\ll 1$, 
and therefore we have neglected the terms non-linear in $\tau$. If we replace the vectors
$u^\mu$ and $n^\mu$ in Eq.(A.14) by the vectors $u'^\mu$ and $n'^\mu$ of Eq.(A.15), we find
that if $T^{\mu\nu}$ is a very slowly varying function of the proper time $\tau$, the rate 
of change of the boost energy through the surface is, when $\tau=0$:
\begin{equation}
\frac{d^2 E_{matter}}{d\tau^2}\vert_{\tau=0} = aA(T_{\mu\nu}u^\mu u^\nu 
+ T_{\mu\nu} n^\mu n^\nu)\vert_{\tau=0}.
\end{equation}
For instance, if our surface is a very small acceleration surface, which is originally at 
rest with respect to our orthonormal geodesic system of coordinates, and is then 
accelerated to the direction of the $z$-axis, the only non-vanishing component of the 
vector $u^\mu$ is
\begin{equation}
u^0 = 1,
\end{equation}
and the only non-vanishing component of the vector $n^\mu$ is
\begin{equation}
n^3 = 1
\end{equation}
at the moment, when $\tau=0$. So we observe from Eq.(A.16) that the rate of change of the
boost energy flow to the direction of the negative $z$-axis is
\begin{equation}
\frac{d^2 E_{matter}}{d\tau^2}\vert_{\tau=0} = aA(T_{00} + T_{33}),
\end{equation}
which, by means of Eqs.(A.3) and (A.4), implies:
\begin{equation}
\frac{d^2 E_{matter}}{d\tau^2}\vert_{\tau=0} = aA(\rho + p_3).
\end{equation}
Since our matter distribution is assumed to be isotropic, we finally get:
\begin{equation}
\frac{d^2 E_{matter}}{d\tau^2}\vert_{\tau=0} = aA(\rho + p).
\end{equation}

   At this point we employ the crucial assumption that our accleration surface is at rest
with respect to our system of coordinates, and therefore also with respect to the matter, 
when $\tau=0$. This means that when $\tau=0$, the boost energy flow through the surface
vanishes. In other words, we have:
\begin{equation}
\frac{dE_{matter}}{d\tau}\vert_{\tau=0} = 0.
\end{equation}
Hence we may write the boost energy flown during a very small proper time interval $\tau$ 
through our acceleration surface as a Taylor expansion:
\begin{equation}
\Delta E_{matter} = \frac{1}{2}\frac{d^2 E_{matter}}{d\tau^2}\vert_{\tau=0}\,\tau^2 
+ O(\tau^3),
\end{equation}
and Eq.(A.21) implies:
\begin{equation}
\Delta E_{matter} = \frac{1}{2}aA(\rho + p)\tau^2 + O(\tau^3).
\end{equation}

   We may now compare the quantity $\Delta E_{matter}$ of Eq.(A.24), which tells the 
amount of boost energy flown through our acceleration surface during a very short proper 
time interval $\tau$, to the quantity $\Delta Q_{as}$ of Eq.(A.13), which tells the heat
change of our accleration surface during that proper time interval. We get, for very small 
$\tau$:
\begin{equation}
\frac{\Delta E_{matter}}{\Delta Q_{as}} = -\frac{3}{2}\frac{\rho + p}{\rho + 3p}.
\end{equation}
Since the pressure $p$ is assumed to be non-negative, and to satisfy Eq.(A.7), the dominant
energy condition, we may write $\Delta E_{matter}$ in the form:
\begin{equation}
\Delta E_{matter} = -\alpha\,\Delta Q_{as},
\end{equation}
where the number $\alpha$ has the property:
\begin{equation}
\frac{3}{4} \leq \alpha \leq \frac{3}{2}.
\end{equation}
So we find that, indeed, $\Delta E_{matter}$ is of the same order of magnitude as 
$-\Delta Q_{as}$, or:
\begin{equation}
\Delta E_{matter} \sim -\Delta Q_{as},
\end{equation}
at least under the assumptions made at the beginning of this Appendix. In other words, we have
obtained Eq.(2.20).

  Of particular interest is the special case, where the spacetime region under consideration
is filled with non-interacting, homogeneous dust. In that case Eq.(A.25) implies:
\begin{equation}
\Delta E_{matter} = -\frac{3}{2}\Delta Q_{as},
\end{equation}
which is Eq.(2.21). Another interesting special case is the one, where matter consists of
massless, non-interacting radiation. In that case it turns out that the boost energy flow 
is exactly the heat flow carried by the radiation, and
\begin{equation}
p = \frac{1}{3} \rho.
\end{equation}
Eq.(A.25) implies:
\begin{equation}
\Delta E_{matter} = -\Delta Q_{as}.
\end{equation}
In other words, the heat of the acceleration surface is exactly converted to the boost 
energy, or heat, of the radiation.

\end{document}